\documentclass[12pt,a4paper]{article}

\usepackage{amssymb, latexsym, amsthm}
\usepackage{graphicx} 
\usepackage{url}      
\newcommand{\doi}[1]{\url{http://dx.doi.org/#1}}
\usepackage{amsmath}  

\newcommand{\p}{\partial}
\newcommand{\D}{\Delta}

\renewcommand{\phi}{\varphi}
\newcommand{\e}{\epsilon}

\newcommand{\R}{{\mathbb R}}

   \newcommand{\PX}{{\Bbb{P}}}
\newcommand{\sde}{\textsc{sde}}
\newcommand{\spde}{\textsc{spde}}

\newtheorem{theorem}{Theorem}
\newtheorem{lemma}[theorem]{Lemma}
\newtheorem{remark}[theorem]{Remark}

\title{Macroscopic reduction for stochastic reaction-diffusion equations}

\author{
W. Wang\thanks{School of Mathematics, University of Adelaide, South
Australia, \textsc{Australia}. \protect\url{mailto:
w.wang@adelaide.edu.au}; and Department of Mathematics, Nanjing
University, Nanjing, \textsc{China}.
\protect\url{mailto:wangweinju@yahoo.com.cn}} \and A.~J.
Roberts\thanks{School of Mathematics, University of Adelaide, South
Australia, \textsc{Australia}.
\protect\url{mailto:anthony.roberts@adelaide.edu.au}} }

\date{\today}

\begin{document}

\maketitle

\begin{abstract}
The macroscopic behavior of dissipative stochastic partial
differential equations usually can be described by a finite
dimensional system. This article proves that a macroscopic reduced
model may be constructed for stochastic reaction-diffusion equations
with cubic nonlinearity by artificial separating the system into two
distinct slow-fast time parts. An averaging method and a deviation
estimate show that the macroscopic reduced model should be a
stochastic ordinary equation which includes the random effect
transmitted from the microscopic timescale due to the nonlinear
interaction. Numerical simulations of an example stochastic heat
equation confirms the predictions of this stochastic modelling
theory.  This theory empowers us to better model the long time
dynamics of complex stochastic systems.
\end{abstract}




\pagestyle{myheadings} \thispagestyle{plain} \markboth{W. Wang \& A.
J. Roberts}{Macroscopic reduction for \textsc{srde}s}

\section{Introduction}\label{sec:intro}
Stochastic partial differential equations (\textsc{spde}s) are
widely studied in modeling, analyzing, simulating and predicting
complex phenomena in many fields of nonlinear
science~\cite[e.g.]{E00, Imkeller, Simulation}.

 Reaction-diffusion equations (\textsc{rde}s) are important mathematic models naturally
 applied in chemistry, biology, geology, physics and ecology etc.
Such equations can be obtained from microscopic particle
 systems under the so called hydrodynamic space-time scaling
 limit~\cite{Chen}\,. When taking fluctuations under the hydrodynamic
 limit into account, an additive noise appears as correction term to the reaction-diffusion
equation~\cite{BPRS}. This reaction-diffusion equation with noise
is viewed as an equation describing intermediate level between the
macroscopic and microscopic ones.  A deterministic equation
appears in the macroscale only if the randomness averages out.

However, for nonlinear complex systems,
random effects may survive averaging and thus be fed into the
macroscopic system~\cite{WR08}. Especially when we consider the
macroscopic behavior of solution on a long  time scale, such random
effects should not be neglected~\cite[e.g.]{Pav07, Rob08, WR08}.
Furthermore, macroscopic turbulence may need to be modeled as noise in, for example,  geophysical fluid dynamics.

Bl\"omker et al.~\cite[e.g.]{BH05,BMP01}
recently studied amplitude equations for \textsc{spde}s with cubic nonlinearity, which is proved
to be a stochastic Landau equation. But in the amplitude equation
the fluctuation from the fast modes disappears when the noise just acts
on fast modes. For \textsc{spde}s with quadratic nonlinearity,
Roberts~\cite{Rob03} derived the amplitude
equation for one dimensional stochastic Burgers equations by
calculating a stochastic normal form model. Bl\"omker et al.~\cite{BHP07} also
gave a rigorous proof for a general \textsc{spde}s with quadratic
nonlinearity by a multiscale analysis; they showed that the
amplitude equations include the fluctuation from the fast mode due
to the nonlinearity interaction.

This paper considers  reaction-diffusion systems driven by a noise
which is homogeneous in space and white in time. Here we are
concerned with the dynamics of the system  on a long time scale. For
this, a scale transformation separates the system into slow and fast
modes. Then we derive a low dimension macroscopic system which
provides the long term dynamics. And the low dimensional macroscopic
system includes a noise term which is transmitted from the fast
modes due to nonlinear interaction.

For definiteness, let the non-dimensional $I=(0, \pi)$ and $L^2(I)$
be the Lebesgue space of square integrable real valued functions
on~$I$\,. Consider the following non-dimensional reaction-diffusion
 equation
 \begin{eqnarray}\label{e:SRD0}
 \p_tw&=&\p_{xx}w+f(w)+\sigma \p_tW\quad
\text{on}\;\; 0<x<\pi\,,\\w&=&0\quad \text{on} \;\;x=0\,,\;
\pi\,,\label{e:SRDbd}
 \end{eqnarray}
 where $f(w)$ represents a nonlinear reaction and $W$ is an $L^2(I)$ valued $Q$-Wiener process defined on complete probability space
$(\Omega, \mathcal{F}, \mathbb{P})$ which is detailed in the next
section. Our aim is to study the behavior of solutions
to~(\ref{e:SRD0})--(\ref{e:SRDbd}) over a large timescale, say
$\e^{-1}$ for small $\e>0$\,. For some fixed integer $N>0$\,, we
split the field~$w$ into $N$~low wavenumber modes and the remaining
high wavenumber modes. For this non-dinesional problem, denote by
$\{e_k(x)\}_k=\{\sin(kx)\}_k$ the orthonormal eigenvectors of
$\p_{xx}$. Define the projection operators onto `slow' and `fast'
modes respectively
\begin{equation}\label{e:PNQN}
\mathcal{P}_N\cdot  =\sum_{k=1}^Ne_k(x)\langle e_k(x)\,, \cdot
\rangle\,,\quad \mathcal{Q}_N=\mathcal{I}-\mathcal{P}_N\,,
\end{equation}
where $\mathcal{I}$ is the identity operator on $L^2(I)$\,. Then
defining $u=\mathcal{P}_Nw$ and $v=\mathcal{Q}_Nw$\,, the
\textsc{rde}~(\ref{e:SRD0}) is identical to the following coupled
equations
\begin{eqnarray}\label{e:split_u}
\p_t u&=&\p_{xx}u+\mathcal{P}_Nf(u+v)+ \sigma \mathcal{P}_N\p_tW\,,
\\
 \p_t v&=&\p_{xx}v+\mathcal{Q}_Nf(u+v)+ \sigma \mathcal{Q}_N\p_tW\,,
 \label{e:split_v}
\end{eqnarray}
whence $w=u+v$\,.
In order to completely separate the time scales we
modify the above system by introducing a `high-pass filter' $A_N$
defined by
\begin{equation*}
A_N=\p_{xx}-(1-\e)\mathcal{P}_N\p_{xx}=(\mathcal{Q}_N+\e
\mathcal{P}_N)\p_{xx}\,.
\end{equation*}
Observed that when $\e=1$\,, the physical case, the operator
$A_N=\p_{xx}$ as appears in~(\ref{e:SRD0}) and
(\ref{e:split_u})--(\ref{e:split_v}); but when $\e=0$\,, the
operator $A_N$ is a pure high-pass filter with null space spanned of
the `slow' modes.

For the moment also assume the noise acts only on the high wavenumber modes, that
is $\mathcal{P}_N W=0$\,. And modify the system~(\ref{e:split_u})--(\ref{e:split_v}) to
\begin{eqnarray}\label{e:ANu}
\p_t u^\e&=&A_Nu^\e+\mathcal{P}_Nf(u^\e+v^\e)\,,
\\
\p_t v^\e&=&A_Nv^\e+\mathcal{Q}_Nf(u^\e+v^\e)+ \sigma\sqrt{\e}
\mathcal{Q}_N\p_tW\,.\label{e:ANv}
\end{eqnarray}
Here the choice of~$\sqrt{\e}$, the factor in front of noise term,
ensures the fast modes, solutions of~(\ref{e:ANv}), remain of order~$1$ as
$t\rightarrow\infty$ and $\e\rightarrow 0$ for any fixed $u^\e$\,.
Note that when $\e=1$\,, (\ref{e:ANu})--(\ref{e:ANv}) is identical
to~(\ref{e:split_u})--(\ref{e:split_v}) and~(\ref{e:SRD0}). We aim
to use analysis based upon small $\e$ to access a useful
approximation at $\e=1$\,.

Section~\ref{sec:highf} proves that, for small enough $\e>0$\,, high
modes~$v^\e(t)$ is approximated by $\sqrt{\e}\eta_*(t)$ over long
timescales ($1/\e$) where $\eta_*$ is the stationary solution
solving the linear stochastic partial differential equation
\begin{equation}\label{e:eta}
\p_t\eta=A_N\eta+\sigma \mathcal{Q}_N\p_tW\,.
\end{equation}
Consequently our careful averaging proves that the macroscopic
behavior of $u^\e(t)$ is described to a first approximation by $\sqrt{\e}u_N(\e t)$ which
solves the following finite dimensional, deterministic system
\begin{equation}\label{e:averaged}
\p_tu_N=\p_{xx}u_N+\mathcal{P}_N\overline{ f_0}(u_N)\,.
\end{equation}
Here, the average
\begin{equation}\label{barf}
\overline{f_0}(u_N)=\lim_{t\rightarrow\infty}
\frac{1}{t}\int_0^tf_0(u_N+\eta_*(s))ds\,,
\end{equation}
and $f_0$ is the cubic component in~$f$, see detail in next section.
Usually~(\ref{e:averaged}) is called the averaged equation and in
probability the following convergence holds (Section~\ref{sec:averaged}): for any $T>0$ there is
positive constant $C$
\begin{equation}\label{e:aveapp}
\sup_{0<t<\e^{-1}T}|u^\e(t)-\sqrt{\e}u_N(\e t)|_{L^2(D)}\leq C\e
\end{equation}
for small~$\e$ under some proper conditions on the initial value.  Further, the martingale approach of Section~\ref{sec:deviation}
shows that small Gaussian fluctuations generally appear in these slow
modes on the timescale~$\e^{-1}$.
The approach shows that fluctuations are transmitted from the fast
modes by nonlinear interactions.  The last Section~\ref{sec:esfom} confirms these theoretical predictions by comparing them to numerical simulations of a specific stochastic reaction diffusion equation.

\section{ Preliminaries and main results}\label{sec:separ}
This section gives some preliminaries and states the main result.
First we give some functional background and some assumptions.

 Let $H=L^2(I)$. Denote
by $A$ the second order operator $\p_{xx}$ with Dirichlet boundary
on $I$ and let $\{e_i\}_{i=1}^\infty$ be a eigen-basis of $H$ such
that
\begin{equation*}
 Ae_i=-\alpha_i e_i\,, \quad i= 1,2,\ldots\,,
\end{equation*}
with $0<\alpha_1<\alpha_2<\alpha_3<\cdots$. For any $\delta>0$\,,
introduce the space $H^{\delta}_0=D(A^{\delta/2})$, which is
compactly embedding into~$H$. And let $H^{-\delta}$~denote the dual
space of~$H_0^\delta$. The usual norm defined on~$H^\delta$ is
written as~$\|\cdot\|_{\delta}$. And for $\delta=0$ and~$1$, the
corresponding norms are written as~$|\cdot|$ and~$\|\cdot\|$
respectively. Denote by $\langle \cdot, \cdot\rangle$ an inner
product in $H$ such as the inner product $\left<u,v\right>=(2/\pi)\int_0^\pi uv\,dx$\,. And for positive integer $N$, denote by $H_N$ the
space spanned by $\{e_1, \ldots, e_N\}$ and by $H_N^\perp$ the space
spanned by $\{e_{N+1}, \ldots \}$\,.

We are given a complete probability space $(\Omega, \mathcal{F},
\{\mathcal{F}_t\}_{t\geq 0}, \mathbb{P})$. Assume $W$ is an
$H$-valued $Q$-Wiener process with operator $Q$ that commutes with
$\mathcal{P}_N$ and satisfies
\begin{equation*}
Qe_i=\lambda_i e_i\,, \quad i= 1,2,\ldots\,,
\end{equation*}
with $\lambda_i=0$\,, $i=1, 2, \ldots\,, N$ and $\lambda_i>0$\,,
$i=N+1, \ldots$\,. Then
\begin{equation}\label{e:wiener}
W(t)=\sum_{i=N+1}^\infty\sqrt{\lambda_i}\beta_i(t)e_i
\end{equation}
 where component noises
$$
\beta_i(t)=\frac{1}{\sqrt{\lambda_i}}\langle W(t), e_i\rangle\,,
\quad i=N+1, \ldots\,,
$$
are real valued Brownian motions mutually independent on $(\Omega,
\mathcal{F}, \mathbb{P})$.
 Moreover we assume
\begin{equation}\label{Q}
\operatorname{tr}Q=\sum_{i=N+1}^\infty\lambda_i<\infty.
\end{equation}
In the following we denote by $Q_2=(I-\mathcal{P}_N)Q=Q$\,. For the
nonlinear term~$f$ we assume the following hypotheses $(\textbf{H})$
\begin{enumerate}
  \item $f:\R\rightarrow \R$ is smooth and has the following form
   $$
   f(\xi)=f_0(\xi)+f_1(\xi)\,,\quad \xi\in\R \,,
   $$
where $f_0(\xi)=-c_0\xi^3$\,, $c_0$ is some positive constant,
$f'''_1(0)=0$ and $f_1(\xi)\xi\leq 0$\,, $\xi\in\R$\,.
   \item   $f'(\xi)<c_1$ for some positive constant $c_1$\,.
  \item There are positive constants $c_2$\,, $c_3$ and positive
  integer $p$ such  that for $\xi\in\R$
  $$
  |f(\xi)|\leq c_2|\xi|^{2p-1}+c_3\,,\quad
  f(\xi)\xi\leq c_2\xi^{2p}+ c_3\,.
  $$
  \item $(f(\xi)-f(\delta))(\xi-\eta)\leq 0\,, \quad \xi, \eta\in\R\,.$
\end{enumerate}
  Then  define
 \begin{equation}\label{e:feps}
 f^\e(\xi)=\frac{1}{\e\sqrt{\e}}f(\sqrt{\e}\xi)
 \end{equation}
which is well-defined as $\e\rightarrow 0$ for any $\xi\in\R$ by the
above assumptions. Then $f^\e$ satisfies the hypotheses (\textbf{H})
with $f_1$ replaced by $f_1^\e(\xi)=f_1(\sqrt{\e}\xi)/\e\sqrt{\e}
$\,.

\begin{remark}
One example for such $f$ is the following polynomial with only
negative odd order terms
$$
f(\xi)=-a_1\xi^3-\cdots-a_p\xi^{2p-1}
$$
where $a_i>0$\,, $i=1, \ldots, p$\,.
\end{remark}

Now we state the main result of this paper.
\begin{theorem} \label{thm:main}
Assume the Wiener process $W$ satisfies assumption~(\ref{Q}). Then for
any time $T>0$ and  constant $R>0$, there is positive constant
$C>0$\,, such that for any solution $(u^\e(t), v^\e(t))$ of~(\ref{e:ANu})--(\ref{e:ANv}) with initial value $|(u_0, v_0)|\leq
\sqrt{\e}R$\,, there is a $N$-dimensional Wiener process
$\bar{W}$ such that  in distribution
\begin{equation}\label{e:u-uN-rho}
\sup_{0\leq t\leq \e^{-1}T}|u^\e(t)-\sqrt{\e}u_N(\e t)-\e\rho_N(\e
t)|\leq C\e^{1+}\,.
\end{equation}
Here $u_N$ solves~(\ref{e:averaged}) and $\rho_N$ solves the
following stochastic differential equations
\begin{equation}\label{rho}
\p_t\rho_N=A_N\rho_N
+\mathcal{P}_N[\overline{f'_0}(u_N)\rho_N]+
\sqrt{B(u_N)} \, \p_t\bar{W}\,,
\end{equation}
with $\rho_N(0)=0$ and
$$
\overline{f'_0}(u_N)=\lim_{t\rightarrow\infty}
\frac{1}{t}\int_0^tf'_0(u_N+ \eta_*(s))ds
$$
and
\begin{eqnarray*}
B(u_N)&=&2\int_0^\infty\mathbb{E}\Big[\mathcal{P}_Nf_0(u_N+\eta_*(t))-\mathcal{P}_N\overline{f_0}(u_N)
\Big]\\&&{}\otimes\Big[\mathcal{P}_Nf_0(u_N+\eta_*(0))-\mathcal{P}_N\overline{f_0}(u_N)\Big]dt\,,
\end{eqnarray*}
where $\otimes$ is the tensor product. Furthermore
\begin{equation}\label{e:v-eta}
\mathbb{E}\sup_{0\leq t\leq
\e^{-1}T}|v^\e(t)-\sqrt{\e}\eta_*(t)|\leq C\e\sqrt{\e}\,.
\end{equation}
\end{theorem}

Now we introduce some scale transformations such that system~(\ref{e:ANu})--(\ref{e:ANv}) is defined on time scale $\e^{-1}$
under the transformations. Introduce slow time $t'=\e t$ and small
fields $u^\e=\sqrt{\e}u'^\e(x, t')$ and $v^\e=\sqrt{\e}v'^\e(x,
t')$\,. Substituting and hereafter omitting primes, the coupled
system~(\ref{e:ANu})--(\ref{e:ANv}) is transformed to the following
stochastic reaction diffusion equation with clearly separated time
scales:
\begin{eqnarray}\label{e:SRDu}
 \p_t u^\e&=&\p_{xx}u^\e+\mathcal{P}_Nf^\e(u^\e+v^\e)\,,
\\
\p_t v^\e&=&\frac{1}{\e}\p_{xx}v^\e+\mathcal{Q}_Nf^\e(u^\e+v^\e)+
\frac{\sigma}{\sqrt{\e}}\mathcal{Q}_N\p_tW \,, \label{e:SRDv}
\end{eqnarray}
with $u^\e(0, t)=u^\e(\pi, t)=v^\e(0, t)=v^\e(\pi, t)=0$ for
$t>0$\,. The Wiener process $W$ is a rescaled version of~(\ref{e:wiener}) and with the same distribution.

For convenience in the following we rewrite~(\ref{e:SRDu})--(\ref{e:SRDv}) into the following form
\begin{eqnarray}\label{e:SRD}
\p_tw^\e=A_\e w^\e+f^\e(w^\e)+\Sigma_\e \p_t{W}\,, \quad
\text{on}\quad  0<x <\pi\,.
\end{eqnarray}
Here $A_\e =(\p_{xx}\,, \e^{-1}\p_{xx})$ with zero Dirichlet
boundary condition.

In order to approximate solutions of~(\ref{e:ANu})--(\ref{e:ANv}),
our basic idea is to pass limit $\e\rightarrow 0$ in~(\ref{e:SRDu})--(\ref{e:SRDv}) to determine interesting structure in
solutions, then study the deviation between the limit and solution.
This first step is to give compact estimates, that is,
tightness of solutions, as addressed in the following two
sections.

\section{Stochastic convolution}\label{sec:conv}

 Start by considering the linear stochastic equation
\begin{equation*}
\p_tz^\e=A_\e z^\e +\Sigma_\e\p_t{W}, \quad z^\e(0)=0 \,.
\end{equation*}
Let $S_\e(t)$ be the analytic semigroup generated by $A_\e$, then in
a mild sense
\begin{equation}\label{e:sconv}
z^\e(t)=\int_0^tS_\e(t-s)\Sigma_\e dW(s).
\end{equation}
For any $T>0$ and $\delta>0$\,, we give a uniform estimates for
$z(t)$, $0<t<T$, in space $H^\delta_0$. We have
\begin{theorem}\label{thm:z-2norm}
 Assume~(\ref{Q}). Then for any $T>0$\,, $q>0$\,,  there is
 positive constant $C_q(T)$ such that
\begin{equation}\label{e:z-2norm}
\mathbb{E}\sup_{0\leq t\leq T}\|z^\e(t)\|^q\leq C_q(T).
\end{equation}
\end{theorem}
\begin{proof}
By using the stochastic factorization formula \cite{PZ92}, for
$\alpha\in (0,1/2)$ we have
\begin{equation*}
z^\e(t)=\frac{\sin\pi\alpha}{\pi}\int_0^t(t-s)^{\alpha-1}S_\e(t-s)Y^\e_\alpha(s)ds
\end{equation*}
where
\begin{equation}\label{e:Ya}
Y_\alpha^\e(s)=\int_0^s(s-r)^{-\alpha}S_\e(s-r)\Sigma_\e \, dW(r).
\end{equation}
Now for $q>1/\alpha$\,, we have by the definition of $A_\e$
\begin{eqnarray*}
\|z^\e(t)\|^q_\delta&\leq&
c_q\Big[\int_0^t(t-s)^{\frac{(\alpha-1)q}{q-1}}
ds\Big]^{q-1}\int_0^t\big\|S_\e(t-s)Y_\alpha^\e(s)\big\|^q_\delta \,
ds\\&\leq &c_q(T)\int_0^t\|Y_\alpha^\e(s)\|^q_\delta\, ds
\end{eqnarray*}
for some positive constant $c_q(T)$. Then we have
\begin{equation*}
\mathbb{E}\sup_{0\leq t\leq T}\|z^\e(t)\|^q_\delta\leq
c_q(T)\mathbb{E}\int_0^T\|Y^\e_\alpha(s)\|^q_\delta\,ds\,.
\end{equation*}
By the definition of $A_\e$,  rewrite~(\ref{e:Ya}) as
\begin{eqnarray*}
Y_\alpha^\e(s)= \sum^\infty_{i=N+1}\sigma\sqrt{\lambda_i}e_i
\int_0^s(s-r)^{-\alpha}e^{-\alpha_i(s-r)/\e}\e^{-1/2}d\beta_i(r).
\end{eqnarray*}
Then if $q>4$, by the Bukholder--Davies--Gundy inequality we have
\begin{eqnarray*}
\mathbb{E}\|Y^\e_\alpha(s)\|^q_\delta&=&\mathbb{E}
\Big\|\int_0^s(s-r)^{-\alpha}S_\e(s-r)\Sigma_\e\, dW(r)\Big\|_\delta^q\\
&\leq &c_q\Big[\sum_{i=N+1}^\infty\sigma^2\lambda_i\alpha_i^\delta
\int_0^s(s-r)^{-2\alpha}e^{-2\alpha_i(s-r)/\e}\e^{-1}\,dr\Big]^{q/2}\\
&\leq &
c_q(s)\sigma^q\Big[\sum_{i=1}^\infty\lambda_i\alpha_i^{\delta-1}\Big]^{q/2}.
\end{eqnarray*}
Therefore there is positive constant $C_q(T)$, independent
of~$\e$, such that
\begin{equation*}
\mathbb{E}\sup_{0\leq t\leq T}\|z^\e(t)\|^q_\delta\leq
c_q(T)\mathbb{E}\int_0^T\|Y^\e_\alpha(s)\|^q_\delta\,ds\leq C_q(T).
\end{equation*}
By the assumption~(\ref{Q}), taking $\delta\leq1$ and the Young
inequality yields the result~(\ref{e:z-2norm}) for all $q>0$\,.
\end{proof}

\section{Tightness of solutions}\label{sec:tight}
This section gives a tightness result by some a priori
estimates for solutions to~(\ref{e:SRDu})--(\ref{e:SRDv}) and the
estimate of Theorem~\ref{thm:z-2norm}.

 Define $\tilde{w}^\e=w^\e-z^\e$, then by~(\ref{e:SRD})
\begin{equation}\label{e: RRD}
 \p_t\tilde{w}^\e=A_\e\tilde{w}^\e+f^\e(w^\e)
\end{equation}
which is equivalent to
\begin{eqnarray*}
\p_t\tilde{u}^\e&=&\p_{xx} \tilde{u}^\e+\mathcal{P}_Nf^\e(w^\e)\,,\\
\p_t \tilde{v}^\e&=&\e^{-1}\p_{xx}
\tilde{v}^\e+\mathcal{Q}_Nf^\e(w^\e)
\end{eqnarray*}
with zero Dirichlet boundary condition on $(0, \pi)$ and
$\tilde{w}^\e=\tilde{u}^\e+\tilde{v}^\e$\,.

 Then we have for some positive constants $c_4$ and $c_5$
\begin{eqnarray*}
\frac{1}{2}\frac{d}{dt}|\tilde{w}^\e|^2&=&
-\|\tilde{u}^\e\|^2-\e^{-1}\|\tilde{v}^\e\|^2+\langle f^\e(w^\e),
\tilde{w}^\e\rangle\\&=&-\|\tilde{u}^\e\|^2-\e^{-1}\|\tilde{v}^\e\|^2+\langle
f^\e(w^\e), w^\e\rangle-\langle f^\e(w^\e), z \rangle\\ &\leq&
-\|\tilde{w}^\e\|^2-c_2|w^\e|^{2p}_{L^{2p}(I)}+c_3\pi+
\left(c_2|w^\e|^{2p-1}_{L^{2p}(I)}+c_3\pi^{1/(2p)'}\right)|z|_{L^{(2p)'}(I)}
\\&\leq& -\|\tilde{w}^\e\|^2-c_4|w^\e|^{2p}_{L^{2p}(I)}+
c_5\left(|z|^{2p}_{L^{(2p)'}(I)}+1\right).
\end{eqnarray*}
Integrating with respect to time yields
\begin{eqnarray}
&&\sup_{0\leq t\leq
T}|\tilde{w}^\e(t)|^2+2\int_0^T\|\tilde{w}^\e(s)\|^2\,ds+2c_4\int_0^T|w^\e(s)|^{2p}_{L^{2p}(I)}\,ds\nonumber
\\&\leq& |w_0|^2+2c_5\int_0^T|z^\e(s)|^{2p}_{L^{(2p)'}(I)}ds+2c_5T\nonumber
 \\&\leq&|w_0|^2+ 2c_5\int_0^T\|z^\e(s)\|^{2p}\,ds+2c_5T.\label{e:w-L2}
\end{eqnarray}

On the other hand we have a positive constant $c_5$ such that
\begin{eqnarray*}
\frac{1}{2}\frac{d}{dt}\|\tilde{w}^\e\|^2&=&-\langle
A_\e\tilde{w}^\e,
\D\tilde{w}^\e\rangle-\langle f^\e(w^\e), \D w^\e+\D z^\e\rangle\\
&\leq
&-|\D\tilde{w}^\e|^2+c_1\|\tilde{w}^\e\|^2+
\left(c_2|w^\e|^{2p-1}_{L^{2p}(I)}+c_3\pi^{1/(2p)'}\right)
\|z^\e\|_{L^{(2p)'}(I)}\\
&\leq&-|\D\tilde{w}^\e|^2+c_1\|\tilde{w}^\e\|^2+
|w^\e|^{2p}_{L^{2p}(I)}+c_6\left(\|z^\e\|^{2p}_{L^{(2p)'}(I)}+1\right)\,.
\end{eqnarray*}
Integrating with respect to time yields
\begin{eqnarray*}
\sup_{0\leq s\leq t}\|\tilde{w}^\e(s)\|^2&\leq&
\|w_0\|^2+2c_1\int_0^t\sup_{0\leq \tau\leq
s}\|\tilde{w}^\e(\tau)\|^2\,ds+2\int_0^T|w^\e(s)|^{2p}_{L^{2p}(I)}\,ds
\\&&{}+ 2c_6\int_0^T\|z^\e(s)\|^{2p}_{L^{(2p)'}(I)}\,ds+2c_6T.
\end{eqnarray*}
Then by the Gronwall lemma and~(\ref{e:w-L2}),
\begin{equation}\label{e:w-H1}
\sup_{0\leq t\leq T}\|\tilde{w}^\e(t)\|^2\leq
c_7T\left(1+\|w_0\|^2+\int_0^T\|z^\e(s)\|^{2p}\,ds\right)
\end{equation}
for some positive constant $c_7$.
 Further for any integer $m\geq 1$\,,
\begin{eqnarray*}
\frac{d}{dt}[\|\tilde{w}^\e\|^2]^m\leq C_m\big[\|\tilde{w}^\e\|^{2m}
+ \|\tilde{w}^\e\|^{2m-2}|w^\e|^{2p}_{L^{2p}(I)}+
\|z^\e\|^{2mp}+1\big].
\end{eqnarray*}
Then
\begin{eqnarray*}
&&\sup_{0\leq s\leq t}\|\tilde{w}^\e(s)\|^{2m}\\&\leq&
C_m\int_0^t\sup_{0\leq \tau\leq
s}\|\tilde{w}^\e(\tau)\|^{2m}\,ds+C_m\sup_{0\leq \tau\leq
T}\|\tilde{w}^\e(\tau)\|^{2m-2}\int_0^T|w^\e(s)|^{2p}_{L^{2p}(I)}\,ds+\\&&
C_m\int_0^T\|z^\e(s)\|^{2mp}\,ds+C_mT+\|w_0\|^{2m}.
\end{eqnarray*}
By induction on $m$, we derive that
\begin{equation}\label{e:w-Hm}
\sup_{0\leq t\leq T}\|\tilde{w}^\e(t)\|^{2m}\leq
C_mT\Big(1+\|w_0\|^{2m}+\int_0^T\|z^\e(t)\|^{2mp}\,dt\Big)
\end{equation}
for some positive constant $C_m$.

Now we show~$\{\mathcal{L}(w^\e)\}_\e$, the distribution of~$w^\e$,
is tight in~$C(0, T; H)$. For this we need the
 following lemma by Simon~\cite{Simon}.

\begin{lemma}\label{lem:compemb}
Assume $E$, $E_0$ and $E_1$ be Banach spaces such that $E_1\Subset
E_0$, the interpolation space $(E_0, E_1)_{\theta,1}\subset E$ with
$\theta\in (0, 1)$  and $E\subset E_0$ with $\subset$ and $\Subset$
denoting continuous and compact embedding respectively. Suppose
$p_0$, $p_1\in [1,\infty]$ and $T>0$, such that
$$
\mathcal{V}\; {\textrm is\; a\;bounded\; set\; in\;} L^{p_1}(0, T;
E_1)
$$
and
$$
\p\mathcal{V}:=\{\p v: v\in \mathcal{V}\}\; {\textrm is\;
a\;bounded\; set\; in\;} L^{p_0}(0, T; E_0).
$$
Here $\p$ denotes the distributional derivative. If
$1-\theta>1/p_\theta$ with
 $$
\frac{1}{p_\theta}=\frac{1-\theta}{p_0}+\frac{\theta}{p_1}\,,
 $$
then $\mathcal{V}$ is relatively compact in $C(0, T; E)$.
\end{lemma}
 Now by the above lemma, noticing the estimate~(\ref{e:z-2norm}) and
 $w^\e=\tilde{w}^\e-z^\e$, we draw the following result
\begin{theorem}
Assume~(\ref{Q}). For any $T>0$\,, $\{\mathcal{L}(w^\e)\}_{\e}$ is
tight in $C(0, T; H)$\,.
\end{theorem}

\section{Macroscopic reduction}\label{sec:reduction}
In this section we prove the main result. First Section~\ref{sec:highf} approximates the high modes  by the Gaussian
process~$\eta^\e$. Then Section~\ref{sec:averaged} derives an averaged approximation for the low modes, and the fluctuation is
considered in Section~\ref{sec:deviation}.

\subsection{Approximation for high modes} \label{sec:highf}
We consider the high frequency dynamics of~(\ref{e:SRDv}). First
for any fixed $u\in H_N$\,, $v^\e$ satisfies
\begin{equation}\label{e:ve}
\p_t
v^\e=\frac{1}{\e}\p_{xx}v^\e+\mathcal{Q}_Nf^\e(u+v^\e)+\frac{\sigma}{\sqrt{\e}}\p_t{W}.
\end{equation}
For any $v_0,v_1\in H_N^\perp$, we have
\begin{equation*}
|v^\e(t; v_0)-v^\e(t; v_1)|^2\leq e^{-2\alpha_{N+1}t/\e}|v_0-v_1|^2
\end{equation*}
which means a unique stationary solution
$\tilde{v}^\e_u$ exists for any fixed~$u$. In the following we  determine
the limit of $\tilde{v}^\e_u$ in $C(0, T; H_N^\perp)$ as
$\e\rightarrow 0$ for any $u\in H_N$\,.

For this we scale time for equation~(\ref{e:eta}) by $t'=\e t$ which
yields, upon omitting primes,
\begin{equation}\label{e:OU}
 \p_t \eta=\frac{1}{\e}\p_{xx}\eta+\frac{\sigma}{\sqrt{\e}}\p_tW.
\end{equation}
Here $W$ is a rescaled version of the noise process in~(\ref{e:eta})
and with the same distribution. Then $\eta^\e_*(t)=\eta^*(t/\e)$ is
the unique stationary solution of~(\ref{e:OU}). Moreover $\eta_*^\e$
is an exponential mixing  Gaussian process with distribution
$\mu=\mathcal{N}\big(0, \sigma^2(-A)^{-1}Q_2/2\big)$.

 Now we prove that for any $u\in H_N$\,, $\tilde{v}_u^\e$ could be
approximated by $\eta^\e$ as $\e$ is small. Let $V^\e=v^\e-\eta^\e$,
then
\begin{equation*}
 \p_tV^\e=\frac{1}{\e}\p_{xx}V^\e+\mathcal{Q}_Nf^\e(u+v^\e),\quad
V^\e(0)=v(0)-\eta^\e(0).
\end{equation*}
Multiplying $V^\e$ on both sides of above equation in $H$ yields
\begin{equation}\label{e:VeL2}
\frac{d}{dt}|V^\e|^2\leq
-\frac{\alpha_{N+1}}{\e}|V^\e|^2+\frac{\e}{2\alpha_{N+1}}|f^\e(u+v^\e)|^2.
\end{equation}
Then by the Gronwall lemma and~(\ref{e:w-Hm}), there is positive
constant $C$ such that for any $t>0$
\begin{eqnarray*}
\mathbb{E}|V^\e(t)|^2&\leq&
e^{-\alpha_{N+1}t/\e}\mathbb{E}|v(0)-\eta_*(0)|^2+
\\&&\quad \frac{\e}{2\alpha_{N+1}}\int_0^te^{-\alpha_{N+1}(t-s)/\e}\mathbb{E}|f^\e(u+v^\e(s))|^2\,ds\\
&\leq&\e^2C\big(\|w(0)\|^{2p}+\mathbb{E}|\eta_*(0)|^2\big).
\end{eqnarray*}
Furthermore by
\begin{equation*}
V^\e(t)=e^{At/\e}
V^\e(0)+\int_0^te^{A(t-s)/\e}\mathcal{Q}_Nf^\e(u+v^\e(s))ds
\end{equation*}
for any $T>0$\,, there is positive constant $C_T$ such that
\begin{equation*}
\mathbb{E}\sup_{0\leq t\leq T}|v^\e(t)-\eta_*^\e(t)|\leq \e C_T\big(
\|w_0\|^{2p}+\mathbb{E}|\eta_*(0)| \big)\,.
\end{equation*}
This proves~(\ref{e:v-eta}).

We end this subsection by giving an estimate on $|V^\e(t)|^{2m}$\,,
$m>0$\,, which is used in the fluctuation estimate in Section~\ref{sec:deviation}. By~(\ref{e:VeL2}) we have
\begin{eqnarray*}
\frac{d}{dt}|V^\e|^{2m}&\leq&
-\frac{m\alpha_{N+1}}{\e}|V^\e|^{2m}+\frac{\e}{2\alpha_{N+1}}|V^\e|^{2m-2}|f(u+v^\e)|^2\\
&\leq& -\frac{m\alpha_{N+1}}{2\e}|V^\e|^{2m}+\e C_m|f(u+v^\e)|^{2m}
\end{eqnarray*}
for some positive constant $C_m>0$\,. Then by the H\"older inequality,
Gronwall lemma and~(\ref{e:w-Hm}) we have
\begin{equation}\label{e:VeL2m}
\mathbb{E}|V^\e(t)|^{2m}\leq C_m\,,\quad t\geq 0\,.
\end{equation}


\subsection{Averaged equation}\label{sec:averaged}
In order to pass limit $\e\rightarrow 0$\,, we restrict our system
into a small probability space. By the estimates in Section~\ref{sec:tight}, for any $\kappa>0$ there is a compact set
$B_\kappa\subset C(0\,, T\,; H)$ such that
$$
\mathbb{P}\{ u^\e\in B_\kappa \}>1-\kappa\,.
$$
Furthermore there is positive constant $C_T^\kappa$, such that
$$
\|u^\e(t)\|^2\leq C_T^\kappa\,,\quad t\in[0, T] ,
$$
for any $u^\e\in B_\kappa$. Now we introduce the
probability space ($\Omega_\kappa\,, \mathcal{F}_\kappa\,,
\mathbb{P}_\kappa$) defined by
$$
\Omega_\kappa=\{\omega\in\Omega: u^\e\in B_\kappa\}\,,\quad
\mathcal{F}_\kappa=\{S\cap \Omega_\kappa: S\in\mathcal{F} \}
$$
and
$$
\mathbb{P}_\kappa(S)=\frac{\mathbb{P}(S\cap\Omega_\kappa)}{\mathbb{P}(\Omega_\kappa)}\,,
\text{ for } S\in\mathcal{F}_\kappa\,.
$$
Then $\mathbb{P}(\Omega\setminus\Omega_\kappa)\leq \kappa$\,.

 Now we restrict $\omega\in\Omega_\kappa$ and introduce an
auxiliary process. For any $T>0$\,, partition the interval~$[0, T]$
into subintervals of length $\delta=\sqrt{\e}$\,. Then we construct
processes~$(\tilde{u}^\e, \tilde{v}^\e)$ such that for $t\in
[k\delta, (k+1)\delta)$,
 \begin{eqnarray}
\tilde{u}^\e(t)&=&e^{A(t-k\delta)}u^\e(k\delta)+
\int_{k\delta}^te^{A(t-s)}\mathcal{P}_Nf^\e(u^\e(k\delta),\tilde{v}^\e(s))\,ds\,, \nonumber\\
&&\tilde{u}^\e(0)=u_0 \,,\label{e:tue}\\
\p_t\tilde{v}^\e(t)&=&\frac{1}{\e}\p_{xx}\tilde{v}^\e(t)+\mathcal{Q}_Nf^\e(u^\e(k\delta),
\tilde{v}^\e(t))+\frac{\sigma}{\sqrt{\epsilon}}\mathcal{Q}_N\,\p_t{W}(t)
\,, \nonumber\\&&
\tilde{v}^\e(k\delta)=v^\e(k\delta)\label{e:tve} \,.
 \end{eqnarray}
Then by the It\^o formula for $t\in [k\delta, (k+1)\delta)$,
\begin{eqnarray*}
&&\frac{1}{2}\frac{d}{dt}|v^\e(t)-\tilde{v}^\e(t)|^2\\&\leq&
-\frac{\lambda_{N+1}}{\e}|v^\e(t)-\tilde{v}^\e(t)|^2+\big\langle
f^\e(u^\e(t)\,,
v^\e(t))-f^\e(u^\e(t)\,, \tilde{v}^\e(t)), v^\e(t)- \tilde{v}^\e(t)\big\rangle\\
&&{}+ \big\langle f^\e(u^\e(t)\,,
\tilde{v}^\e(t))-f^\e(u^\e(k\delta)\,, \tilde{v}^\e(t))\,, v^\e(t)-
\tilde{v}^\e(t)\big\rangle\\&\leq &
-\frac{\lambda_{N+1}}{2\e}|v^\e(t)-\tilde{v}^\e(t)|^2+\e\left(c_2\|w^\e(t)\|^{2p}+c_3\right)|u^\e(t)-\tilde{u}^\e(k\delta)|^2\,.
\end{eqnarray*}
By the choice of $\Omega_\kappa$, there is $C_T>0$\,, such that
\begin{equation}\label{e:u1}
|u^\e(t)-u^\e(k\delta)|^2\leq C_T\delta^2\,, \quad \text{for} \;t\in
[k\delta, (k+1)\delta)\,.
\end{equation}
Then by the Gronwall lemma,
\begin{equation}\label{e:v1}
|v^\e(t)-\tilde{v}^\e(t)|^2\leq C_T\delta^2\,,\quad t\in [0, T]\,.
\end{equation}
In a mild sense for $t\in [k\delta, (k+1)\delta)$
\begin{equation*}
u^\e(t)=e^{A(t-k\delta)}u^\e(k\delta)+
\int_{k\delta}^te^{A(t-s)}\mathcal{P}_Nf^\e(u^\e(s),v^\e(s))\,ds\,.
\end{equation*}
Then by the cubic property of $f$ and smoothing property of $e^{A
t}$, noticing the choice of $\Omega_\kappa$, we have for
$t\in[k\delta, (k+1)\delta)$
\begin{eqnarray*}
|u^\e(t)-\tilde{u}^\e(t)|&\leq&
 C'\int_{k\delta}^t|v^\e(s)-\tilde{v}^\e(s)|\,ds+
C'\int_{k\delta}^t|u^\e(k\delta)-u^\e(s)|\,ds
\end{eqnarray*}
for some positive constant $C'$\,. So by~(\ref{e:v1}) we have
\begin{equation}\label{e:u2}
|u^\e(t)-\tilde{u}^\e(t)|\leq C_T\delta\,,\quad  t\in[0\,, T]\,.
\end{equation}

On the other hand, in a mild sense the solution of~(\ref{e:averaged})
 is
\begin{equation*}
u_N(t)=e^{At}u_0+\int_0^te^{A(t-s)}\mathcal{P}_N\overline{f_0}(u_N(s))\,ds\,.
\end{equation*}
Then, using $\lfloor z\rfloor$~to denote the largest integer less
than or equal to~$z$,
\begin{eqnarray*}
|\tilde{u}^\e(t)-u_N(t)|&\leq&\int_0^te^{A(t-s)}\big|\mathcal{P}_Nf^\e(u^\e(\lfloor
s/\delta\rfloor\delta),
\tilde{v}^\e(s))-\mathcal{P}_N\overline{f^\e}(u^\e(\lfloor
s/\delta\rfloor\delta))\big|ds \\
&&{}+\int_0^te^{A(t-s)}\big|\mathcal{P}_N\overline{f^\e}(u^\e(\lfloor
s/\delta\rfloor\delta)
)-\mathcal{P}_N\overline{f_0}(u^\e(s))\big|ds\\
&&{}+\int_0^te^{A(t-s)}\big|\mathcal{P}_N\overline{f_0}(u^\e(s))-\mathcal{P}_N\overline{f_0}(u_N(s))\big|ds
\,.
\end{eqnarray*}
Notice $\eta$ is independent of $\e$, by the assumption
\textbf{H},  and $\mathcal{P}_N\overline{f^\e}$ is continuous in~$\e$.  Moreover the exponential mixing stationary measure~$\mu$
is independent of~$u$, by the ergodic theorem
\begin{equation*}
\overline{
f_0}(u_N)=\lim_{t\rightarrow\infty}\frac{1}{t}\int_0^tf_0(u_N+
\eta_*(s))ds=\int_{H_N^\perp}f_0(u_N+v)\mu(dv)\,.
\end{equation*}
Then for any $u_1$\,, $u_2\in H_N$
\begin{eqnarray}
&& \Big|\int_{H_N^\bot}[f_0(u_1,
v)-f_0(u_2,
 v)\big]\mu(dv)\Big|\nonumber\\
 &\leq &
 2c_0\Big|\int_{H_N^\bot}(u_1-u_2)(u^2_1+u_2^2+v^2)\mu(dv)\Big|
 \nonumber \\
 &\leq & 2c_0\Big[\|u_1\|^2+\|u_2\|^2+\mathbb{E}\|\eta^*\|^2
 \Big]|u_1-u_2|\,\label{e:PNf-conti}
\end{eqnarray}
which yields the continuity of $\mathcal{P}_N\overline{f}_0\,:
H_N\rightarrow H_N$\,. Then we have for $t\in [0, T]$
\begin{equation}\label{e:u3}
|\tilde{u}^\e(t)-u_N(t)|\leq
C_T\left[\delta+\int_0^T|u^\e(s)-u_N(s)|\,ds\right] \,.
\end{equation}
As
\begin{equation*}
|u^\e(t)-u_N(t)|\leq |u^\e(t)-\tilde{u}(t)|+|\tilde{u}(t)-u_N(t)| \,,
\end{equation*}
by the Gronwall lemma and~(\ref{e:u1}), (\ref{e:u2})
and~(\ref{e:u3}) we have for $t\in [0, T]$,
\begin{equation}\label{e:u0}
|u^\e(t)-u_N(t)|\leq C_T\sqrt{\e} \,.
\end{equation}

Now by the arbitrariness of~$\kappa$,  we complete the proof of the
averaging approximation. And since $\eta_*$ is
Gaussian with zero mean, we give
\begin{equation}\label{e:barf}
\mathcal{P}_N\overline{
f_0}(u_N)=-c_0\mathcal{P}_N(u_N^3+3u_N\mathbb{E}\bar{\eta}^2)\,.
\end{equation}

\subsection{Fluctuation}\label{sec:deviation}
This subsection  details the approximation of $u^\e$ for small~$\e$.
We study the deviation between $u^\e$ and $u$, which proves to be a
Gaussian process. This shows that there are  fluctuations in the
slow modes. We follow an approach used previously~\cite{WR08,Kes79,
Wata88}. For this define the scaled difference
\begin{equation*}
\bar{\rho}^\e=\frac{1}{\sqrt{\e}}(u^\e-u_N)\,.
\end{equation*}
Then $\bar{\rho}^\e$ solves
\begin{equation*}
\p_t\bar{\rho}^\e=\p_{xx}\bar{\rho}^\e+\frac{1}{\sqrt{\e}}
\big[\mathcal{P}_N f^\e(u^\e,
v^\e)-\mathcal{P}_N\overline{f^\e}(u_N)\big]\,, \quad
\bar{\rho}^\e(0)=0\,.
\end{equation*}
However, here we just consider $\rho^\e$, the solution of
\begin{equation*}
\p_t\rho^\e=\p_{xx}\rho^\e+\frac{1}{\sqrt{\e}} \big[\mathcal{P}_N
f_0(u^\e, v^\e)-\mathcal{P}_N\overline{f_0}(u_N)\big]\,, \quad
\rho^\e(0)=0.
\end{equation*}
By assumption (\textbf{H}) and estimate~(\ref{e:w-Hm})\,,
$\mathbb{E}|\bar{\rho}^\e(t)-\rho^\e(t)|\rightarrow 0$ as
$\e\rightarrow 0$ for any $t\geq 0$.\\

Noticing the estimate~(\ref{e:VeL2m}), by the Gronwall lemma for
any $T>0$\,,
\begin{equation}\label{e:Ez}
\mathbb{E}\sup_{0\leq t\leq
T}|\rho^\e(t)|^2+\mathbb{E}\int_0^T\|\rho^\e(t)\|^2dt\leq
C_T(1+\|w_0\|^6) \,.
\end{equation}
In the mild sense we write
 \begin{displaymath}
\rho^\e(t)=\frac{1}{\sqrt{\e}}\int_0^te^{A(t-r)}
\left[\mathcal{P}_Nf_0(u^\e(r),
v^\e(r))-\mathcal{P}_N\overline{f_0}(u_N(r))\right]dr.
 \end{displaymath}
Then for any $0\leq s<t$\,, by the property of~$e^{At}$, we have for
some positive $1>\delta>0$
\begin{eqnarray*}
|\rho^\e(t)-\rho^\e(s)|&\leq&
\frac{1}{\sqrt{\e}}\Big|\int_0^te^{A(t-r)}
\mathcal{P}_N f_0(u^\e(r), v^\e(r))-\mathcal{P}_N\overline{f_0}(u_N(r) )dr\\
&&{}-\int_0^se^{A(s-r)}
\mathcal{P}_Nf_0(u^\e(r), v^\e(r))-\mathcal{P}_N\overline{f_0}(u_N(r) )dr\Big|\\
&\leq& C_T|t-s|^\delta\frac{1}{\sqrt{\e}}\big|\mathcal{P}_Nf_0(u^\e,
v^\e)-\mathcal{P}_N\overline{f_0}(u_N)\big|_{L^2(0, T; H_N)} \,.
\end{eqnarray*}
By~(\ref{e:Ez}) and the estimates in Section~\ref{sec:tight}
\begin{displaymath}
\mathbb{E}\frac{1}{\sqrt{\e}}\big|\mathcal{P}_Nf_0(u^\e,
v^\e)-\mathcal{P}_N\overline{f_0}(u_N)\big|_{L^2(0, T; H_N)}\leq
C_T(1+\|w_0\|^6).
\end{displaymath}
Then
\begin{equation}\label{e:zCtheta}
\mathbb{E}|\rho^\e(t)|_{C^\delta(0, T; H_N)}\leq C_T(1+\|w_0\|^6)
\,.
\end{equation}
Here $C^\delta(0, T; H_N)$ is the H\"older space  with
exponent~$\delta$.
 On the other hand, also by the property
of~$e^{At}$, we have for some positive constant~$C_{T, \alpha}$ and
for some $1>\alpha>0$
\begin{eqnarray*}
|\rho^\e(t)|_{H^\alpha}&\leq&\frac{1}{\sqrt{\e}}\int_0^t(t-s)^{-\alpha/2}
\big|\mathcal{P}_Nf_0(u^\e(s), v^\e(s))-\mathcal{P}_N\overline{f_0}(u_N(s))\big|ds \nonumber\\
& \leq&  C_{T, \alpha}\frac{1}{\sqrt{\e}}\big|\mathcal{P}_Nf_0(u^\e,
v^\e)-\mathcal{P}_N\overline{f_0}(u_N)\big|_{L^2(0, T; H_N)} \,.
\end{eqnarray*}
Then
\begin{equation}\label{e:zCtheta2}
\mathbb{E}\sup_{0\leq t\leq T}|\rho^\e(t)|_{H_N^\alpha}\leq C_{T,
\alpha}(1+\|w_0\|^6) \,.
\end{equation}
And by the compact embedding of $C^\delta(0, T; H_N)\cap C(0, T;
H_N^\alpha)\subset C(0, T; H_N)$, $\{\nu^\e\}_\e$, the distribution
of~$\{\rho^\e\}_\e$ is tight in~$C(0, T; H)$.

Split $\rho^\e=\rho_1^\e+\rho_2^\e$ where each component satisfies, respectively,
\begin{eqnarray*}&&
\p_t\rho_1^\e=\p_{xx}\rho_1^\e+\frac{1}{\sqrt{\e}}\big[\mathcal{P}_Nf_0(u_N,
\eta^\e)-\mathcal{P}_N\overline{f_0}(u_N)\big]\,, \quad
\rho_1^\e(0)=0 \,,
\\&&
\p_t\rho_2^\e=\p_{xx}\rho_2^\e+\frac{1}{\sqrt{\e}}\big[\mathcal{P}_Nf_0(u^\e,
v^\e)-\mathcal{P}_Nf_0(u_N, \eta^\e)\big]\,, \quad \rho_2^\e(0)=0 \,.
\end{eqnarray*}

Denote by~$\nu_1^\e$ the probability measure of~$\rho_1^\e$ induced
on space~$C(0,T; H_N)$.  And for $\gamma>0$ denoted
by~$UC^\gamma(H_N, \R)$ the space of all functions from~$H_N$
to~$\R$ which are uniformly continuous on~$H_N$ together with all
Fr\'echet derivatives to order~$\gamma$. In the following, for any
$h\in UC^\gamma(H_N, \R)$ and $u_N\in H_N$\,, denote by $\langle
h'(u_N), \cdot\rangle: H_N\rightarrow \R$ the linear map defined by
the first order Fr\'echet derivatives of $h$ and $h''(u_N)(\cdot):
H_N\otimes H_N\rightarrow \R$ the linear map defined by the second
order Fr\'echet derivatives of $h$\,. Then we have
\begin{lemma}\label{lem:mart}
 Any limiting measure of~$\nu_1^\e$, denote by~$P^0$, solves the
following martingale problem on~$C(0, T; H_N)$:
$P^0\{\rho_1(0)=0\}=1$\,,
\begin{displaymath}
h(\rho_1(t))-h(\rho_1(0))-\int_0^t \langle h'(\rho_1(\tau)),
A\rho_1(\tau)\rangle d\tau-\frac{1}{2}\int_0^t
 \operatorname{tr}\big[h''(\rho_1(\tau))(B(u_N))\big ] d\tau
\end{displaymath}
is a $P^0$-martingale for any $h\in UC^2(H_N, \R)$. Here
\begin{eqnarray*}
B(u_N)&=&2\int_0^\infty \mathbb{E} \left[(\mathcal{P}_Nf_0(u_N,
\bar{\eta}(t))-\mathcal{P}_N\overline{f_0}(u_N))
\right.\\&&\left.\quad{}
\otimes
(\mathcal{P}_Nf_0(u_N,
\bar{\eta}(0))-\mathcal{P}_N\overline{f_0}(u_N))\right]dt.
\end{eqnarray*}
\end{lemma}
\begin{proof}
 For any $0<s\leq t<\infty$ and $h\in UC^\infty(H)$ we
have
\begin{eqnarray*}
&&h(\rho_1^\e(t))-h(\rho_1^\e(s))\\&=& \int_s^t\left\langle
h'(\rho_1^\e(\tau)),\frac{d\rho_1^\e}{dt}\right\rangle\, d\tau
\\&=&
\int_s^t\left\langle h'(\rho_1^\e(\tau)),
A\rho_1^\e(\tau)\right\rangle\,
d\tau
\\&&{}+\frac{1}{\sqrt{\e}}\int_s^t\left\langle
h'(\rho_1^\e(\tau)), \mathcal{P}_Nf_0(u_N(\tau),
\eta^\e(\tau))-\mathcal{P}_N\overline{f_0}(u_N(\tau))
\right\rangle\, d\tau \,.
\end{eqnarray*}
Rewrite the second term as
 \begin{eqnarray*}
&&\frac{1}{\sqrt{\e}}\int_s^t\left\langle h'(\rho_1^\e(\tau)),
\mathcal{P}_Nf_0(u_N(\tau),
\eta^\e(\tau))-\mathcal{P}_N\overline{f_0}(u_N(\tau))\right\rangle \, d\tau \\
&=& \frac{1}{\sqrt{\e}}\int_s^t\left\langle h'(\rho_1^\e(t)),
\mathcal{P}_Nf_0(u_N(\tau),
\eta^\e(\tau))-\mathcal{P}_N\overline{ f_0}(u_N(\tau))\right\rangle\, d\tau\\
&&{}-\frac{1}{\sqrt{\e}}\int_s^t\!\!\!\int_\tau^t
h''(\rho_1^\e(\delta))\Big(\mathcal{P}_Nf_0(u_N(\tau),
\eta^\e(\tau))-\mathcal{P}_N\overline{f_0}(u_N(\tau)))
\\&&\qquad
\otimes A\rho_1^\e(\delta)\Big)\, d\delta \,d\tau\\
&&{}-\frac{1}{\e}\int_s^t\!\!\!\int_\tau^t
h''(\rho_1^\e(\delta))\Big( \mathcal{P}_Nf_0(u_N(\tau),
\eta^\e(\tau))-\mathcal{P}_N\overline{f_0}(u_N(\tau))
\\&&{}\qquad \otimes \mathcal{P}_Nf_0(u_N(\delta),
\eta^\e(\delta))-\mathcal{P}_N\overline{f_0}(u_N(\delta)) \Big) \,
d\delta\, d\tau\\&=&L_1+L_2+L_3 \,.
\end{eqnarray*}
Let $\{e_i\}_{i=1}^\infty$ be one eigenbasis of~$H$, then
\begin{eqnarray*}
&&h''(\rho_1^\e(\delta))\Big( \mathcal{P}_Nf_0(u_N(\tau),
\eta^\e(\tau))-\mathcal{P}_N\overline{f_0}(u_N(\tau))\otimes
\\&&\qquad
\mathcal{P}_Nf_0(u_N(\delta),
\eta^\e(\delta))-\mathcal{P}_N\overline{f_0}(u_N(\delta)) \Big)
\\&&
=\sum^N_{i,j=1}\p_{ij}h(\rho^\e_1(\delta))\Big\langle
\{\mathcal{P}_Nf_0(u_N(\tau),
\eta^\e(\tau))-\mathcal{P}_N\overline{f_0}(u_N(\delta))\}
\\ &&\qquad{}
\otimes \{\mathcal{P}_Nf_0(u_N(\delta),
\eta^\e(\delta))-\mathcal{P}_N\overline{f_0}(u_N)\}, e_i\otimes e_j
\Big\rangle \,.
\end{eqnarray*}
Here $\p_{ij}=\p_{e_i}\p_{e_j}$ where $\p_{e_i}$~is the directional
derivative in direction~$e_i$.

Denote by
\begin{eqnarray*}
A^\e_{ij}(\delta, \tau)&=&\big\langle
\{\mathcal{P}_Nf_0(u_N(\tau),
\eta^\e(\tau))-\mathcal{P}_N\overline{f_0}(u_N(\tau))\}
\\&&\quad{}
\otimes
\{\mathcal{P}_Nf_0(u_N(\delta),\eta^\e(\delta))-\mathcal{P}_N\overline{f_0}(u_N(\delta))\},
e_i\otimes e_j\big\rangle .
\end{eqnarray*}
Then we have
\begin{eqnarray*}
 L_3&=&-\frac{1}{\e}\sum_{ij}\int_s^t\!\!\!\int_\tau^t
 \p_{ij}h(\rho_1^\e(\delta))\langle A^\e(\delta,\tau)e_i,e_j \rangle \, d\delta\, d\tau\\
&=&-\frac{1}{\e}\sum_{ij}\int_s^t\!\!\!\int_\tau^t\!\!\!
\int_\delta^t\big\langle \p_{ij}h'(\rho_1^\e(\lambda)),
\\&&\qquad{}
A\rho_1^\e(\lambda)+\frac{1}{\sqrt{\e}}\big[\mathcal{P}_Nf_0(u_N(\lambda),
\eta^\e(\lambda))-\mathcal{P}_N\overline{f_0}(u_N(\lambda))\big]\big\rangle
\\&&\qquad{}\times \tilde{A}^\e_{ij}(\delta,\tau)\, d\lambda\, d\delta\, d\tau
\\
\end{eqnarray*}
\begin{eqnarray*}
&&{}
 +\frac{1}{\e}\sum_{ij}\int_s^t\!\!\!\int_\tau^t\p_{ij}h(\rho_1^\e(t))\tilde{A}^\e_{ij}
(\delta,\tau)\, d\delta\,
d\tau\\&&{}+\frac{1}{\e}\sum_{ij}\int_s^t\!\!\!\int_s^\tau\p_{ij}h(\rho_1^\e(\tau))
\mathbb{E}[A^\e_{ij}(\delta,\tau)]\, d\delta \, d\tau\\
&=&L_{31}+L_{32}+L_{33}
\end{eqnarray*}
where  $\tilde{A}_{ij}^\e(\delta, \tau)=A^\e_{ij}(\delta,
\tau)-\mathbb{E}[A^\e_{ij}(\delta, \tau)]$. For our purpose, for any
bounded continuous function~$\Phi$ on~$C(0,s; H)$, let $\Phi(\cdot,
\omega)=\Phi(\rho_1^\e(\cdot, \omega))$.
  Then by the exponential mixing of $\eta^\e$,
\begin{eqnarray*}
|\mathbb{E}[(L_{31}+L_{32})\Phi]|\rightarrow 0  \text{ as }
\e\rightarrow 0 \,.
\end{eqnarray*}
Now we determine the limit of
$\int_s^\tau\mathbb{E}A^\e_{ij}(\delta, \tau)\,d\delta$ as
$\e\rightarrow 0$\,. For this introduce
\begin{eqnarray*}
\bar{A}_{ij}^\e(\delta,
\tau)&=&\Big\langle\left\{\mathcal{P}_Nf_0(u(\tau),
\eta^\e(\tau))-\mathcal{P}_N\overline{f_0}(u(\tau))\right\}
\\&&\quad
\otimes\left\{\mathcal{P}_Nf_0(u(\tau),
\eta^\e(\delta))-\mathcal{P}_N\overline{f_0}(u(\tau))\right\}, e_i\otimes
e_j\Big\rangle\,.
\end{eqnarray*}
Then
\begin{eqnarray*}
&&\Big|\int_s^\tau \mathbb{E}\big[A^\e_{ij}(\delta,
\tau)-\bar{A}^\e_{ij}(\delta,
\tau)\big]\,d\delta\Big|\\&\leq&\int_s^\tau\Big|\mathbb{E}\Big[\langle
\mathcal{P}_Nf_0(u(\tau),
\eta^\e(\tau))-\mathcal{P}_N\overline{f_0}(u(\tau)), e_i\rangle
\\&&{}\quad\times\langle\mathcal{P}_Nf_0(u(\delta), \eta^\e(\delta))-
\mathcal{P}_Nf_0(u(\tau), \eta^\e(\delta)), e_j\rangle\Big]
\Big|d\delta
\end{eqnarray*}
By the assumption $\textbf{H}$ we have
\begin{eqnarray*}
&&\big|\langle \mathcal{P}_Nf_0(u(\delta),
\eta^\e(\delta))-\mathcal{P}_Nf_0(u(\tau), \eta^\e(\delta)),
e_j\rangle\big|\\&\leq& 2\big[\|u(\delta)\|^2+\|u(\tau)\|^2
\big]|u(\delta)-u(\tau)| |e_j| \,,
\end{eqnarray*}
and by~(\ref{e:PNf-conti})
\begin{eqnarray*}
&&\big|\langle
\mathcal{P}_N\overline{f_0}(u(\tau))-\mathcal{P}_N\overline{f_0}(u(\delta)),
 e_j\rangle\big|\\&\leq& 2\Big[\|u(\delta)\|^2+\|u(\tau)\|^2
 +\mathbb{E}\|\eta\|^2\Big]|u(\delta)-u(\tau)| |e_j|\,.
\end{eqnarray*}
Then also by the exponential mixing property of $\eta^\e$
\begin{equation}
\frac{1}{\e}\Big|\int_s^\tau \mathbb{E}[A^\e_{ij}(\delta,
\tau)-\bar{A}^\e_{ij}(\delta, \tau)]\,d\delta\Big|\rightarrow
0\,, \quad \e\rightarrow 0\,.
\end{equation}
Now we put
\begin{eqnarray*}
b^{ij}_{u_N}(\delta-\tau
)&=&\mathbb{E}\big[\big\langle \{\mathcal{P}_Nf_0(u_N,
\eta(\delta))-\mathcal{P}_N\overline{f_0}(u_N)\}
\\&&\qquad{}
\otimes\{ \mathcal{P}_Nf_0(u_N, \eta(\tau))-\mathcal{P}_N\overline{f_0}(u_N)\},
e_i\otimes e_j\big\rangle \big] \,.
\end{eqnarray*}
Then
\begin{equation*}
\mathbb{E}\big[A^\e_{ij}(\delta,
\tau)\big]=b^{ij}_{u_N}\Big(\frac{\delta-\tau}{\e}\Big) \,.
\end{equation*}
Further, by the exponential mixing property, for any fixed
$\delta>\tau$
\begin{equation*}
\int_0^{(\delta-\tau)/\e}b^{ij}_{u_N}(\lambda) \,
d\lambda\rightarrow \int_0^\infty b^{ij}_{u_N}(\lambda) \,
d\lambda=:\frac{1}{2}B_{ij}(u_N)\,, \quad \e\rightarrow 0 \,.
\end{equation*}
Then, if $\e_n\rightarrow 0$ as $n\rightarrow\infty$\,,
$\nu^{\e_n}\rightarrow P^0 $\,,
\begin{equation*}
\lim_{n\rightarrow\infty}\mathbb{E}[L_3\Phi] =\frac{1}{2}
\int_s^t\mathbb{E}^{P^0}\Big(
\operatorname{tr}\big[h''(\rho_1(\tau))B(u_N)\big]\Phi\Big) d\tau
\,,
\end{equation*}
with $B(u_N)=\sum_{ij}B_{ij}(u_N)e_i\otimes e_j$\,. Similarly by the
exponential mixing of $\eta^\e$
\begin{equation*}
\mathbb{E}[L_1\Phi+L_2\Phi]\rightarrow 0 \text{ as }  \e\rightarrow
0 \,.
\end{equation*}
By the tightness of~$\rho^\e$ in~$C(0, T; H)$, the
sequence~$\rho_1^{\e_n}$ has a limit process, denote by~$\rho_1$, in the weak sense. Then
\begin{equation*}
\lim_{n\rightarrow\infty}\mathbb{E} \Big[\int_s^t\langle
h'(\rho_1^{\e_n}(\tau)), A\rho_1^{\e_n}(\tau)\rangle\Phi \,
d\tau\Big]=\mathbb{E}\Big[\int_s^t \langle h'(\rho_1(\tau)),
A\rho_1(\tau)\rangle \Phi \, d\tau \Big]
\end{equation*}
and
\begin{equation*}
\lim_{n\rightarrow\infty}\mathbb{E}\big[\big(h(\rho_1^{\e_n}(t))-
h(\rho_1^{\e_n}(s))\big)\Phi\big]=\mathbb{E}\big[\big(h(\rho_1(t))-h(\rho_1(s))\big)\Phi\big]
\,.
\end{equation*}
At last we have
\begin{eqnarray}&&
\mathbb{E}^{P^0}\big[\big(h(\rho_1)(t)-h(\rho_1(s))\big)\Phi\big]
\nonumber\\
&=&\mathbb{E}^{P^0}\Big[\int_s^t\langle h'(\rho_1(\tau)),
A\rho_1(\tau)\rangle \Phi \, d\tau
\Big]
\nonumber\\&&{}
+\frac{1}{2}\mathbb{E}^{P^0}\left\{\int_s^t \operatorname{tr}
\big[h''(\rho_1(\tau))B(u_N)\big]\Phi \, d\tau \right\}.
\label{e:z1mart}
\end{eqnarray}
By an approximation argument we prove~(\ref{e:z1mart}) holds for all
$h\in UC^2(H)$. This completes the proof.
\end{proof}

By~(\ref{e:barf}) we have a more explicit expression of $B(u_N)$ as
\begin{eqnarray}\label{e:BuN}
B(u_N)&=&2\mathbb{E}\int_0^\infty
 \big[\mathcal{P}_N(3u_N(\eta^2-\mathbb{E}\eta^2)+3u^2_N\eta+\eta^3) \big]\\&&
\quad{}\otimes
\big[\mathcal{P}_N(3u_N(\eta^2(0)-\mathbb{E}\eta^2)+3u^2_N\eta(0)+\eta^3(0))
\big]\,dt\,.\nonumber
\end{eqnarray}

Then by the relation between weak solution to \textsc{spde}s and the martingale problem \cite{MM88}, $P^0$ uniquely solves the
martingale problem related to the following stochastic differential
equation
\begin{equation}\label{e:limit-z1}
\p_t\rho_1=A\rho_1 + \sqrt{B(u_N)} \, \p_t\bar{W} \,, \quad \rho_1(0)=0 \,,
\end{equation}
where $\bar{W}(t)$ is $N$-dimensional standard Wiener process,
defined on a probability space~$(\bar{\Omega}, \bar{\mathcal{F}},
\bar{\PX})$ such that $\rho^\e_1$~converges weakly to~$\rho_1$
in~$C(0, T; H_N)$.

 By earlier results~\cite{WR08}, $\rho_2^\e$~converges weakly
to $\rho_2$ in~$C(0, T; H_N)$ and $\rho_2$ uniquely solves
\begin{equation}\label{e:limit-z2}
\p_t{\rho}_2=A\rho_2+\mathcal{P}_N[\overline{ f'_0}(u_N)(\rho_1+\rho_2)]
\,, \quad \rho_2(0)=0 \,.
\end{equation}
Furthermore by $f_0(u_N)=-c_0u_N^3$\,, we have
$\overline{ f'_0}(u_N)=-3c_0\mathbb{E}\bar{\eta}^2$\,.
Then $\rho^\e$ converges weakly in~$C(0, T; H_N)$ to $\rho_N$ which
uniquely solves the following $N$-dimensional stochastic
differential equation
\begin{equation*}
\p_t\rho_N=A\rho_N-3c_0\mathcal{P}_N\left[u_N^2\rho_N+(\mathbb{E}\bar{\eta}^2)\rho_N\right]\,+\sqrt{B(u_N)}
\, \p_t\bar{W} \,, \quad \rho_N(0)=0\,.
\end{equation*}

\section{Example of stochastic force in one mode}
\label{sec:esfom}
This section applies the previous results to a simple case to see one example of
how the noise forcing of high modes feeds into the dynamics of low
modes. Further, we compare the result with that of the stochastic slow manifold model. For simplicity we assume the stochastic force acts just on
the second spatial mode~$\sin 2x$.
\begin{figure}
\centering
\includegraphics{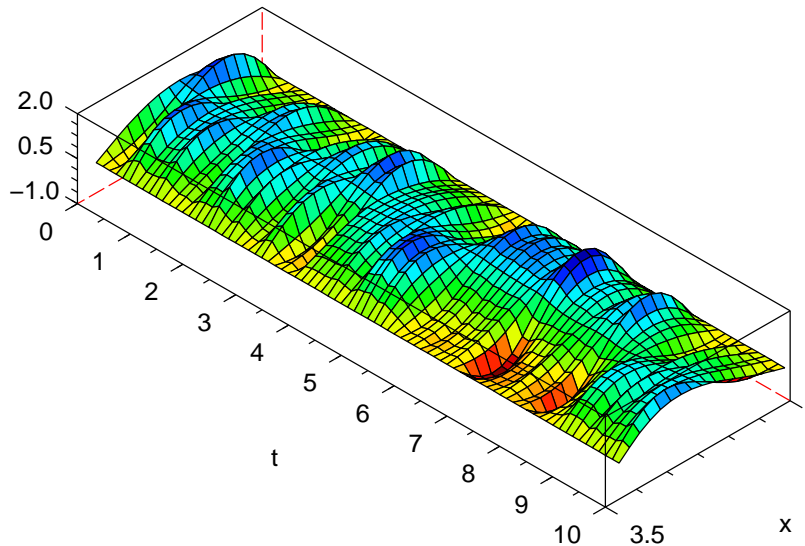}
\caption{one realisation of the space-time dependence of the
stochastic field~$w(x,t)$ for parameters
$\epsilon\gamma=\sqrt\epsilon\sigma=1$\,.  The stochastically forced
$\sin 2x$ mode interacts nonlinearly with the finite amplitude fundamental
mode $\sin x$\,.  The numerics are based on finite differences with $15$~points in space.} \label{fig:stoRD}
\end{figure}

\subsection{Averaging and deviation}

Consider the following stochastic forced heat equation on the domain
 $[0\,, \pi]$
\begin{eqnarray}
 \p_tw=\p_{xx}w+(1+\e\gamma)w-w^3+\sigma\sqrt{\e} \p_tW
 \label{eq:spde1}
\end{eqnarray}
with $w(0, t)=w(\pi, t)=0$\,.  $\gamma$ is a real bifurcation parameter. The spatiotemporal noise~$W$ is defined by~(\ref{e:wiener}) with $\lambda_2=1$ and $\lambda_i=0$
for $i\neq 2$\,, that is, $W(x,t)=\beta_2(t)\sin 2x$\,.    Then only the second spatial mode is forced by white noise. Figure~\ref{fig:stoRD} plots one realisation illustrating the nonlinear dynamics induced by the noise of strength~$\sqrt\e\sigma$.  In the \spde~\eqref{eq:spde1} we incorporate the growth linear in~$w$ to counteract the dissipation on a finite domain so that we can control the clarity of the separation between fast and slow modes. We take $A_1=\p_{xx}+1$ with Dirichlet boundary condition on $[0, \pi]$, then the eigenmodes $e_i(x)=\sin(ix)$ corresponding to decay rates $\alpha_i=i^2-1$\,: giving the slow mode $\sin x$\,; and the fast modes $\sin ix$ for $i\geq 2$\,.  As the parameter~$\epsilon\gamma$ crosses zero with no noise,~$\sigma=0$, there is a deterministic bifurcation to a finite amplitude of the fundamental mode~$\sin x$\,.

We consider the stochastic system~\eqref{eq:spde1} on long timescales of order $\e^{-1}$\,.
First notice that here $f(w)=\e w-w^3$\,, but all the analysis in Section~\ref{sec:reduction} holds. Then decompose the field $w^\e(t)=\sqrt{\e}u^\e(t')+\sqrt{\e}v^\e(t')$ in the slow time $t'=\e t$\,.  By Theorem~\ref{thm:main}
the fast mode~$v^\e$ is approximated by a stationary process~$\eta$ which
solves the following linear equation for small $\e>0$
\begin{equation*}
\p_{t'}\eta=\frac{1}{\e}\p_{xx}\eta+\frac{\sigma}{\sqrt{\e}}Q_N\p_{t'}W\,.
\end{equation*}
Decomposing $\eta=\sum_i\eta_ie_i$, then $\eta_i=0$ for $i\neq 2$ and
the scalar stationary process~$\eta_2$ then satisfies the following stochastic
ordinary differential equation
\begin{equation*}
d\eta_2=-\frac{3}{\e}\eta_2 \,dt'+ \sigma
\sqrt{\frac{1}{\e}}\,d\beta'_2\,.
\end{equation*}
The distribution of $\eta_2$ is the one dimensional
normal distribution~$\mathcal{N}\big(0\,,\frac16{\sigma^2}
\big)$.

\begin{figure}
\centering
\begin{tabular}{c@{}c}
\rotatebox{90}{\hspace{20ex}$\bar a^2$}&
\includegraphics{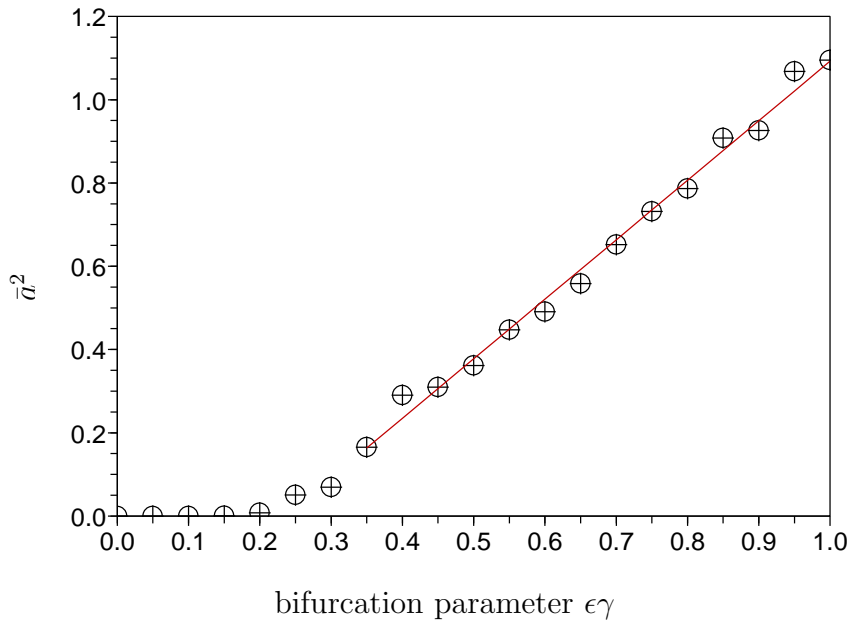}\\
&bifurcation parameter $\epsilon\gamma$
\end{tabular}
\caption{average equilibrium amplitude (squared) of the fundamental mode,~$\bar a^2$, versus bifurcation parameter~$\epsilon\gamma$ for fixed noise amplitude $\sqrt\epsilon\sigma=1$\,.  The straight line fit is $\bar a^2\approx+1.41\epsilon\gamma -0.32$\,.}
\label{fig:stoRDsamps1}
\end{figure}

Theorem~\ref{thm:main} also asserts that the averaged equation for~$u^\e$ is
\begin{equation}\label{e:AvSHE}
\p_{t'}u=\gamma
u-\mathcal{P}_1u^3-\frac{3\sigma^2}{6}\mathcal{P}_1
u\sin^22x\,.
\end{equation}
Suppose $u=A(t')\sin x$, by~(\ref{e:AvSHE}) the amplitude~$A$ satisfies the Landau equation
\begin{equation}\label{e:Av-a}
\frac{dA}{dt'}=\Big(\gamma-\frac{\sigma^2}{4}\Big)A-\frac{3}{4}A^3\,.
\end{equation}
Figure~\ref{fig:stoRDsamps1} plots the average amplitudes,~$\bar a$, of the fundamental mode obtained from numerical simulations.  The figure shows that the amplitude of the fundamental is depressed by the noise in the second component causing a delay in the bifurcation in the presence of noise.  More extensive numerical simulations suggest that the mean amplitude depends upon noise and bifurcation parameter (for $0\leq\epsilon\gamma,\epsilon\sigma^2<1$) according to
\begin{equation}
\bar a^2\approx\epsilon\big(1.32\,\gamma-0.34\,\sigma^2\big)
+\epsilon^2\big(0.08\,\gamma^2+0.03\,\gamma\sigma^2 +0.02\,\sigma^4\big).
\label{eq:fitnum}
\end{equation}
The analytic Landau equation predicts equilibrium amplitude $A^2=\frac43\gamma-\frac13\sigma^2$ which (after scaling by~$\epsilon$) agrees remarkably well with these numerical estimates.

\begin{figure}
\centering
\begin{tabular}{c@{}c}
\rotatebox{90}{\hspace{20ex}$a$}&
\includegraphics{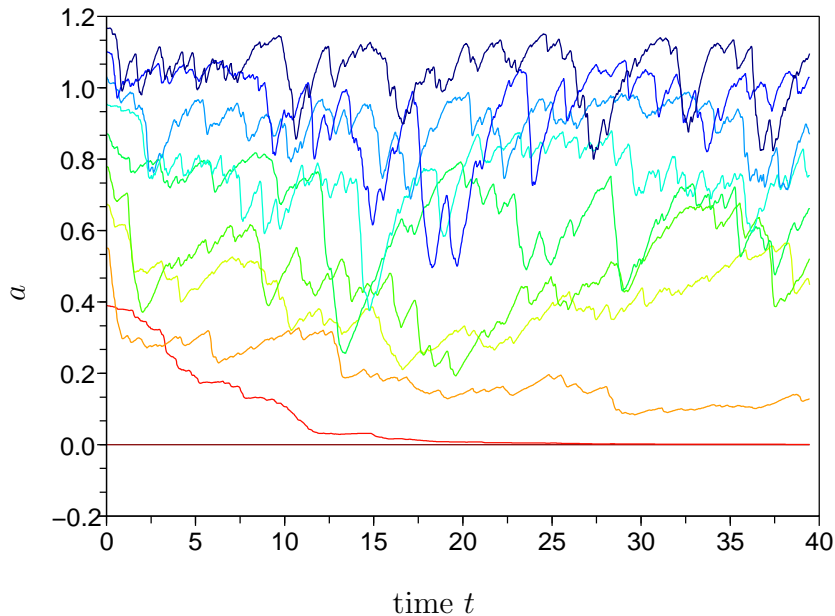}\\
&time~$t$
\end{tabular}
\caption{realisations of the amplitude of the fundamental mode~$a(t)$ versus time for various bifurcation parameters~$\epsilon\gamma$ and for noise amplitude $\sqrt\epsilon\sigma=1$\,.  At time zero the fundamental modes are started at the \emph{deterministic} stable equilibrium for the corresponding~$\epsilon\gamma$.}
\label{fig:stoRDs}
\end{figure}

\begin{figure}
\centering
\begin{tabular}{c@{\ }c}
\rotatebox{90}{\hspace{10ex}standard deviation $\sigma_a$}&
\includegraphics{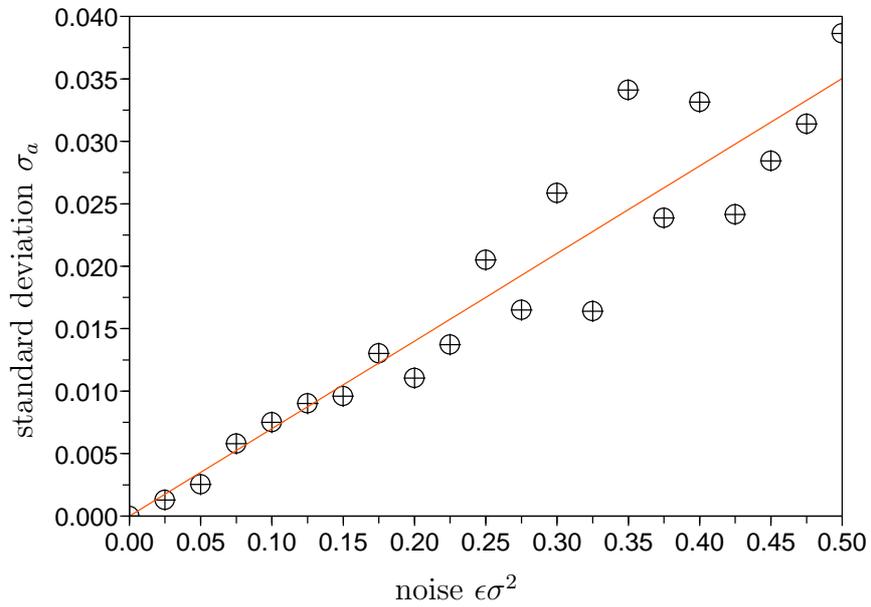}\\[-1ex]
&noise $\epsilon\sigma^2$
\end{tabular}
\caption{standard deviation of the amplitude~$a(t)$ of the fundamental mode~$\sin x$ versus noise~$\epsilon\sigma^2$ for fixed bifurcation parameter $\epsilon\gamma=1$\,.  These are estimated from long time versions of the simulations shown in Figure~\ref{fig:stoRDs}.  The straight line fit predicts $\sigma_a\approx0.07\epsilon\sigma^2$.}
\label{fig:stoRDsdev}
\end{figure}

However, there is also a significant stochastic component in the fundamental mode, $\sin x$\,, as shown by Figure~\ref{fig:stoRDs} which plots the amplitude of the fundamental as a function of time for various bifurcation parameters~$\epsilon\gamma$.  The stochastic fluctuations come from nonlinear interactions of the noise in the $\sin 2x$ mode.
Now we calculate this deviation.  Noting that
$\eta=\eta_2\sin 2x$\,, Lemma~\ref{lem:mart} asserts
\begin{eqnarray*}
 B(A)&=&2\mathbb{E}\int_0^\infty\big[\mathcal{P}_1(u+\eta(s))^3-\overline{\mathcal{P}_1(u+\eta)^3}
\big]\big[\mathcal{P}_1(u+\eta(0))^3-\overline{\mathcal{P}_1(u+\eta)^3} \big]\,ds\\
&=&18A^2\int_0^\infty\mathbb{E}\Big[\big(\eta^2_2(s)-\mathbb{E}\eta^2_2\big)
\big(\eta^2_2(0)-\mathbb{E}\eta^2_2\big)\Big]\langle
e_1^2, e^2_2 \rangle^2\, ds\\
&=&\frac{\sigma^4}{24}A^2\,.
\end{eqnarray*}
Then writing $\lim_{\e\to 0}(u^\e-u)/{\sqrt{\e}}:=\rho_1 \sin x$\,, by
$3\mathcal{P}_1[(\mathbb{E}\eta^2)\rho_1]=\sigma^2\rho_1/4$\,,
the deviation~$\rho_1$ solves the Ornstein--Uhlenbeck-like \sde
\begin{equation}\label{e:devaition}
d\rho_1=\left(\gamma-\frac{\sigma^2}{4}-\frac{9}{4}A^2\right)\rho_1\,dt+\frac{\sigma^2}{2\sqrt{6}}A\,d\beta\,,\quad
\rho_1(0)=0\,,
\end{equation}
where $\beta$~is a standard real valued Brownian motion.  For example, after the mean amplitude~$A$ reaches equilibrium, $A=\sqrt{4(\gamma-\sigma^2/4)/3}$, the \sde~\eqref{e:devaition} predicts fluctuations in~$\rho_1$ with a standard deviation
\begin{equation}
  \sigma_1\approx\frac{\sigma^2}{6\sqrt2}= 0.1179\,\epsilon\sigma^2.
  \label{eq:egfluc}
\end{equation}
Such fluctuations are seen in the numerical simulations of Figure~\ref{fig:stoRDs}.  More extensive numerical simulations estimates the standard deviation of the fluctuations; Figure~\ref{fig:stoRDsdev} plots this standard deviation against the noise amplitude.\footnote{For larger stochastic forcing, $\epsilon\sigma^2>0.5$ at this bifurcation parameter~$\epsilon\gamma$, the standard deviation~$\sigma_a$ appears to plateau.}  A straight line fit to this data gives the standard deviation of the numerically observed fluctuations as $\sigma_a\approx 0.08\,\sigma^2$.  The theoretical prediction~\eqref{eq:egfluc} scales the same with applied noise~$\sigma$, although the coefficient is about~$30\%$ different, and is similarly independent of the bifurcation parameter~$\epsilon\gamma$.  Averaging and deviation together reasonably predict the dynamics of this example \spde~\eqref{eq:spde1}.

\subsection{Compare with the stochastic slow manifold}

Earlier work constructing stochastic slow manifolds of dissipative \spde{}s~\cite{Roberts05c} is easily adapted to the example \spde~\eqref{eq:spde1}.  Recall we choose  $W=\beta_2(t)\sin 2x$\,.  In terms of the amplitude~$a(t)$ of the fundamental mode~$\sin x$\,, computer algebra readily derives that the stochastic slow manifold of the \spde~\eqref{eq:spde1} is
\begin{eqnarray}
w&=&a\sin x+\frac1{32}a^3\sin 3x
+\sqrt\epsilon\sigma\sin 2x\,e^{-3t}\star\dot\beta_2
\nonumber\\&&{}
+\epsilon^{3/2}\gamma\sigma\sin2x\,e^{-3t}\star e^{-3t}\star\dot\beta_2
+\cdots\,.
\end{eqnarray}
The history convolutions of the noise, $e^{-3t}\star\dot\beta_2=\int_{-\infty}^t e^{-3(t-s)}d\beta_2(s)$ that appear in the shape of this stochastic slow manifold empower us to eliminate such history integrals in the evolution \emph{except} in the nonlinear interactions between noises; here simply
\begin{displaymath}
\dot a=\epsilon\gamma a-\frac34a^3
-\frac12\epsilon\sigma^2a\,\big(\dot\beta_2e^{-3t}\star\dot\beta_2\big)+\cdots\,.
\end{displaymath}
Analogously to the averaging and devaition theorems, analysis of Fokker--Planck equations~\cite{Chao95,Roberts05c} then asserts that the cannonical quadratic noise interaction term in this equation should be replaced by the sum of a mean drift and an effectively new independent noise process.
Thus the evolution on this stochastic slow manifold is
\begin{equation}
da\approx \left[\epsilon\left(\gamma-\frac14\sigma^2\right)a-\frac34a^3\right]dt
+\frac1{2\sqrt6}\epsilon\sigma^2a\,d\tilde \beta +\cdots\,,
\label{eq:ssmm}
\end{equation}
where $\tilde \beta$ is a real valued standard Brownian motion.
The stochastic model~\eqref{eq:ssmm} is exactly the averaged equation~(\ref{e:Av-a}) plus the deviation~(\ref{e:devaition}) with $a(t)=\sqrt{\epsilon}A(\e t)+\e\rho_1(\e t)$\,.
And the stochastic model~(\ref{eq:fitnum}) predicts a stochastic equilibrium
amplitude squared of about $\bar
a^2=\frac43\epsilon\gamma-\frac13\epsilon\sigma^2$ in agreement  with the empirical
fit~\eqref{eq:fitnum} to the numerical data of
Figure~\ref{fig:stoRDsamps1} and other simulations.

Now investigate the fluctuations about the stochastic equilibrium as seen in Figure~\ref{fig:stoRDs} and measured in Figure~\ref{fig:stoRDsdev}.  As for the deviation equation~\eqref{e:devaition}, the
stochastic slow model~\eqref{eq:ssmm} is approximately an
Orstein--Uhlenbeck process in the vicinity of the finite amplitude
stochastic equilibrium.  Without elaborating the details, the
form of~\eqref{eq:ssmm} then predicts  fluctuations about the
equilibrium have a standard deviation of~$\epsilon\sigma^2/(6\sqrt2)$ in agreement with~\eqref{eq:egfluc} and in moderate agreement with the fit of Figure~\ref{fig:stoRDsdev} to the numerical simulations.   The stochastic slow manifold model also predicts the stochastic dynamics of this example \spde~\eqref{eq:spde1}.  The difference is that the stochastic slow manifold model is encapsulated in the one \sde~\eqref{eq:ssmm} instead of being split into separate equations for the average and deviation.

\paragraph{Acknowledgements} This research is supported by the
Australian Research Council grant DP0774311 and NSFC grant 10701072.

\end{document}